\documentclass[conference]{IEEEtran}
\IEEEoverridecommandlockouts
\usepackage{cite}
\usepackage{amsmath,amssymb,amsfonts}
\usepackage{algorithmic}
\usepackage{graphicx}
\usepackage{textcomp}
\usepackage{xcolor}
\usepackage{balance}
\def\BibTeX{{\rm B\kern-.05em{\sc i\kern-.025em b}\kern-.08em
    T\kern-.1667em\lower.7ex\hbox{E}\kern-.125emX}}

\newtheorem{assumption}{Assumption}
\newtheorem{theorem}{Theorem}
\newtheorem{lemma}{Lemma}

\newcommand{\aeq}{\overset{\mathrm{a.s.}}{=}}
\newcommand{\ato}{\overset{\mathrm{a.s.}}{\to}}

\begin{document}

\title{On the Convergence of Orthogonal/Vector AMP: 
Long-Memory Message-Passing Strategy
\thanks{The author was in part supported by the Grant-in-Aid 
for Scientific Research~(B) (JSPS KAKENHI Grant Numbers 21H01326), Japan.}
}

\author{
\IEEEauthorblockN{Keigo Takeuchi}
\IEEEauthorblockA{Dept.\ Electrical and Electronic Information Eng., 
Toyohashi University of Technology, 
Toyohashi 441-8580, Japan\\
Email: takeuchi@ee.tut.ac.jp} 
}

\maketitle

\begin{abstract}
This paper proves the convergence of Bayes-optimal orthogonal/vector 
approximate message-passing (AMP) to a fixed point in the large system 
limit. The proof is based on Bayes-optimal long-memory (LM) message-passing 
(MP) that is guaranteed to converge systematically. The dynamics of 
Bayes-optimal LM-MP is analyzed via an existing state evolution framework. 
The obtained state evolution recursions are proved to converge. 
The convergence of Bayes-optimal orthogonal/vector AMP is proved by confirming 
an exact reduction of the state evolution recursions to those for 
Bayes-optimal orthogonal/vector AMP. 
\end{abstract}


\section{Introduction}
Consider the problem of reconstructing an $N$-dimensional sparse signal 
vector $\boldsymbol{x}\in\mathbb{R}^{N}$ from compressed, noisy, and linear 
measurements $\boldsymbol{y}\in\mathbb{R}^{M}$~\cite{Donoho06,Candes061} 
with $M\leq N$, given by 
\begin{equation} \label{model}
\boldsymbol{y} = \boldsymbol{A}\boldsymbol{x} + \boldsymbol{w}. 
\end{equation}
In (\ref{model}), $\boldsymbol{w}\sim
\mathcal{N}(\boldsymbol{0},\sigma^{2}\boldsymbol{I}_{M})$ denotes an additive 
white Gaussian noise (AWGN) vector with variance $\sigma^{2}>0$. The matrix 
$\boldsymbol{A}\in\mathbb{R}^{M\times N}$ represents a known sensing matrix. 
The signal vector $\boldsymbol{x}$ has zero-mean independent and identically 
distributed (i.i.d.) elements with unit variance.  
The triple $\{\boldsymbol{A}, \boldsymbol{x}, \boldsymbol{w}\}$ is 
assumed to be independent random variables. 

A promising approach to efficient reconstruction is message passing (MP), 
such as approximate MP (AMP)~\cite{Donoho09,Rangan11} and 
orthogonal/vector AMP (OAMP/VAMP)~\cite{Ma17,Rangan192}. When the sensing 
matrix has zero-mean i.i.d.\ sub-Gaussian elements, 
AMP was proved to be Bayes-optimal in the large system 
limit~\cite{Bayati11,Bayati15}, where both $M$ and $N$ tend to infinity 
with the compression ratio $\delta=M/N\in(0, 1]$ kept constant. 
However, AMP fails to converge for non-i.i.d.\ cases, such as the 
non-zero mean case~\cite{Caltagirone14} and the ill-conditioned 
case~\cite{Rangan191}.  

OAMP~\cite{Ma17} or equivalently VAMP~\cite{Rangan192} is a powerful MP 
algorithm to solve this convergence issue for AMP. In this paper, these MP 
algorithms are referred to as OAMP. When the sensing matrix is 
right-orthogonally invariant, OAMP was proved to be Bayes-optimal in the 
large system limit~\cite{Rangan192,Takeuchi201}. 

Strictly speaking, the Bayes-optimality of OAMP requires an implicit assumption 
under which state evolution recursions for OAMP converge to a fixed point  
after an infinite number of iterations~\cite{Rangan192,Takeuchi201}. Thus, 
this assumption needs to be confirmed for individual 
problems~\cite{Liu16,Gerbelot20,Mondelli21}. The purpose of this paper 
is to prove this assumption for OAMP using the Bayes-optimal 
denoiser---called Bayes-optimal OAMP.   

The proof is based on a Bayes-optimal long-memory (LM) MP algorithm that is 
guaranteed to converge systematically. LM-MP uses messages in all preceding 
iterations to update the current message while conventional MP utilizes 
messages only in the latest iteration. LM-MP was originally proposed via 
non-rigorous dynamical functional theory~\cite{Opper16,Cakmak17} and 
formulated via rigorous state evolution~\cite{Takeuchi211}. A unified 
framework in \cite{Takeuchi211} was used to propose convolutional 
AMP~\cite{Takeuchi202,Takeuchi211}, memory AMP~\cite{Liu21}, and 
VAMP with warm-started conjugate gradient (WS-CG)~\cite{Skuratovs20}. 
See \cite{Fan22,Venkataramanan21} for another state evolution approach.  

A first step in the proof is a formulation of Bayes-optimal LM-OAMP, in which 
a message in the latest iteration is regarded as an additional measurement 
that depends on all preceding messages. Thus, use of the additional 
measurement never degrades the reconstruction performance if statistical 
properties of the dependent messages are grasped completely and if the current 
message is updated in a Bayes-optimal manner. 

A second step is an application of the unified framework in \cite{Takeuchi211} 
to state evolution analysis for LM-OAMP. It is sufficient to 
confirm that a general error model in \cite{Takeuchi211} contains an error 
model for LM-OAMP. The obtained state evolution recursions represent 
asymptotic correlation structures for all messages in LM-OAMP. Furthermore, 
asymptotic Gaussianity for estimation errors~\cite{Takeuchi211} implies that 
the correlation structures provide full information on the asymptotic 
distributions of the estimation errors. As a result, it is possible 
to update the current message in a Bayes-optimal manner. 

A third step is to prove that state evolution recursions for Bayes-optimal 
LM-OAMP converge to a fixed point under mild assumptions. While the 
convergence is intuitively expected from the formulation of Bayes-optimal 
LM-OAMP, a rigorous proof is non-trivial and based on a novel statistical 
interpretation for optimized LM damping in \cite{Liu21}. 

The last step is an exact reduction of state evolution recursions for 
Bayes-optimal LM-OAMP to those for conventional Bayes-optimal 
OAMP~\cite{Ma17}. Thus, the convergence of Bayes-optimal LM-OAMP 
implies the convergence of conventional Bayes-optimal OAMP to a fixed point. 
As a by-product, the LM-MP proof strategy in this paper claims that 
conventional Bayes-optimal OAMP is the best in terms of convergence speed 
among all possible LM-MP algorithms included into a unified framework in 
\cite{Takeuchi211}. 

The remainder of this paper is organized as follows: Section~\ref{sec2} 
reviews Bayes-optimal estimation based on dependent Gaussian measurements. 
The obtained results reveal a statistical interpretation for optimized 
LM damping~\cite{Liu21}, which is utilized to formulate LM-OAMP in 
Section~\ref{sec3}. 
Section~\ref{sec4} presents state evolution analysis of LM-OAMP 
via a unified framework in \cite{Takeuchi211}. Two-dimensional (2D) discrete 
systems---called state evolution recursions---are derived to describe the 
asymptotic dynamics of LM-OAMP. Furthermore, we prove the convergence 
and reduction of the state evolution recursions. See \cite{Takeuchi213} 
for the details of the proof. 

Finally, see a recent paper~\cite{Liu212} for an application of the LM-MP 
strategy in this paper. 

\section{Correlated AWGN Measurements} \label{sec2}
This section presents a background to define the Bayes-optimal denoiser in 
LM-OAMP. We review Bayesian estimation of a scalar signal $X\in\mathbb{R}$ 
from $t+1$ correlated AWGN measurements $\boldsymbol{Y}_{t}
=(Y_{0},\ldots,Y_{t})\in\mathbb{R}^{1\times(t+1)}$, given by 
\begin{equation} \label{AWGN_measurement}
\boldsymbol{Y}_{t} = X\boldsymbol{1}^{\mathrm{T}} + \boldsymbol{W}_{t}.  
\end{equation}
In (\ref{AWGN_measurement}), $\boldsymbol{1}$ denotes a column vector whose 
elements are all one. The signal $X$ follows the same distribution as that of 
each element in the i.i.d.\ signal vector $\boldsymbol{x}$. 
The AWGN vector $\boldsymbol{W}_{t}\sim\mathcal{N}(\boldsymbol{0},
\boldsymbol{\Sigma}_{t})$ is a zero-mean Gaussian 
row vector with covariance $\boldsymbol{\Sigma}_{t}$ and independent of $X$. 
The covariance matrix $\boldsymbol{\Sigma}_{t}$ is assumed to be positive 
definite.  

This paper uses a two-step approach in computing the posterior mean estimator 
of $X$ given $\boldsymbol{Y}_{t}$. A first step is computation of a sufficient 
statistic $S_{t}\in\mathbb{R}$ for estimation of $X$ given $\boldsymbol{Y}_{t}$. 
The second step is evaluation of the posterior mean estimator of $X$ given 
the sufficient statistic $S_{t}$. This two-step approach is useful in proving 
Lemma~\ref{lemma1} while it is equivalent to direct computation of the 
posterior mean estimator of $X$ given $\boldsymbol{Y}_{t}$, i.e.\ 
$\mathbb{E}[X|S_{t}]=\mathbb{E}[X|\boldsymbol{Y}_{t}]$. 

As shown in \cite[Section~II]{Takeuchi213}, a sufficient statistic for 
estimation of $X$ is given by 
\begin{equation} \label{sufficient_statistic}
S_{t} = \frac{\boldsymbol{Y}_{t}\boldsymbol{\Sigma}_{t}^{-1}\boldsymbol{1}}
{\boldsymbol{1}^{\mathrm{T}}\boldsymbol{\Sigma}_{t}^{-1}\boldsymbol{1}}
= X + \tilde{W}_{t}, \quad \tilde{W}_{t} 
= \frac{\boldsymbol{W}_{t}\boldsymbol{\Sigma}_{t}^{-1}\boldsymbol{1}}
{\boldsymbol{1}^{\mathrm{T}}\boldsymbol{\Sigma}_{t}^{-1}\boldsymbol{1}},
\end{equation}
where $\{\tilde{W}_{\tau}\}$ are zero-mean Gaussian random variables 
with covariance 
\begin{IEEEeqnarray}{rl}
&\mathbb{E}[\tilde{W}_{t'}\tilde{W}_{t}]
=\frac{\boldsymbol{1}^{\mathrm{T}}\boldsymbol{\Sigma}_{t'}^{-1}
\mathbb{E}[\boldsymbol{W}_{t'}^{\mathrm{H}}\boldsymbol{W}_{t}]
\boldsymbol{\Sigma}_{t}^{-1}\boldsymbol{1}}
{\boldsymbol{1}^{\mathrm{T}}\boldsymbol{\Sigma}_{t'}^{-1}\boldsymbol{1}
\boldsymbol{1}^{\mathrm{T}}\boldsymbol{\Sigma}_{t}^{-1}\boldsymbol{1}} 
\nonumber \\ 
=& \frac{\boldsymbol{1}^{\mathrm{T}}\boldsymbol{\Sigma}_{t'}^{-1}
(\boldsymbol{I}_{t'}, \boldsymbol{O})\boldsymbol{\Sigma}_{t}
\boldsymbol{\Sigma}_{t}^{-1}\boldsymbol{1}}
{\boldsymbol{1}^{\mathrm{T}}\boldsymbol{\Sigma}_{t'}^{-1}\boldsymbol{1}
\boldsymbol{1}^{\mathrm{T}}\boldsymbol{\Sigma}_{t}^{-1}\boldsymbol{1}} 
= \frac{1}{\boldsymbol{1}^{\mathrm{T}}\boldsymbol{\Sigma}_{t}^{-1}\boldsymbol{1}}
\label{sufficient_covariance}
\end{IEEEeqnarray}
for all $t'\leq t$. An important observation is that the 
covariance~(\ref{sufficient_covariance}) is independent of $t'$ as long as 
$t'$ is smaller than or equal to $t$. This is a key property in proving 
the reduction of Bayes-optimal LM-OAMP to conventional Bayes-optimal OAMP. 

The Bayes-optimal estimator is defined as the 
posterior mean $f_{\mathrm{opt}}(S_{t};\mathbb{E}[\tilde{W}_{t}^{2}])
=\mathbb{E}[X|S_{t}]$ of $X$ given the sufficient statistic $S_{t}$. 
The posterior covariance of $X$ given $S_{t'}$ and $S_{t}$ is given by 
\begin{IEEEeqnarray}{rl}
C(S_{t'},S_{t};\mathbb{E}[\tilde{W}_{t'}^{2}], \mathbb{E}&[\tilde{W}_{t}^{2}]) 
= \mathbb{E}\left[
 \{X - f_{\mathrm{opt}}(S_{t'};\mathbb{E}[\tilde{W}_{t'}^{2}])\}
\right. \nonumber \\
&\left.
 \left.
  \cdot\{X - f_{\mathrm{opt}}(S_{t};\mathbb{E}[\tilde{W}_{t}^{2}])\}
 \right| S_{t'}, S_{t} 
\right]. \label{posterior_covariance} 
\end{IEEEeqnarray} 
Note that the posterior covariance depends on the noise covariance 
$\mathbb{E}[\tilde{W}_{t'}\tilde{W}_{t}]$, which is not presented explicitly, 
because of $\mathbb{E}[\tilde{W}_{t'}\tilde{W}_{t}]
=\mathbb{E}[\tilde{W}_{t}^{2}]$ for $t'\leq t$.  
These definitions are used to define the Bayes-optimal denoiser. 

We prove key technical results to prove the convergence of state evolution 
recursions for Bayes-optimal LM-OAMP and its reduction to those for 
Bayes-optimal OAMP.  

\begin{lemma} \label{lemma1}
\begin{itemize}
\item $\mathbb{E}[\tilde{W}_{t'}^{2}]\geq\mathbb{E}[\tilde{W}_{t}^{2}]$ and 
$\mathbb{E}[ 
\{X - f_{\mathrm{opt}}(S_{t'};\mathbb{E}[\tilde{W}_{t'}^{2}])\}^{2}]
\geq\mathbb{E}[\{X - f_{\mathrm{opt}}(S_{t};\mathbb{E}[\tilde{W}_{t}^{2}])\}^{2}]$ 
hold for all $t'<t$. 

\item $C(S_{t'},S_{t};\mathbb{E}[\tilde{W}_{t'}^{2}], 
\mathbb{E}[\tilde{W}_{t}^{2}]) = C(S_{t},S_{t};\mathbb{E}[\tilde{W}_{t}^{2}], 
\mathbb{E}[\tilde{W}_{t}^{2}])$ holds for all $t'<t$. 

\item If $[\boldsymbol{\Sigma}_{t}]_{\tau',\tau}
=[\boldsymbol{\Sigma}_{t}]_{\tau,\tau'}
=[\boldsymbol{\Sigma}_{t}]_{\tau,\tau}$ holds for all $\tau'<\tau$, then  
$\mathbb{E}[\tilde{W}_{t}^{2}]=[\boldsymbol{\Sigma}_{t}]_{t,t}$ holds.
\end{itemize}
\end{lemma}
\begin{IEEEproof}
We prove the first property. Since $\{Y_{\tau}\}_{\tau=0}^{t'}\subset
\{Y_{\tau}\}_{\tau=0}^{t}$ holds for all $t'<t$, the optimality of the posterior 
mean estimator implies $\mathbb{E}[ 
\{X - f_{\mathrm{opt}}(S_{t'};\mathbb{E}[\tilde{W}_{t'}^{2}])\}^{2}]
\geq\mathbb{E}[\{X - f_{\mathrm{opt}}(S_{t};\mathbb{E}[\tilde{W}_{t}^{2}])\}^{2}]$. 
The other monotonicity $\mathbb{E}[\tilde{W}_{t'}^{2}]
\geq\mathbb{E}[\tilde{W}_{t}^{2}]$ follows 
from the monotonicity of the MSE with respect to the variance.   

We next prove the second property. Since (\ref{sufficient_covariance}) 
implies $\mathbb{E}[\tilde{W}_{t'}\tilde{W}_{t}]=\mathbb{E}[\tilde{W}_{t}^{2}]$ 
for $t'<t$, the sufficient statistic $S_{t'}$ can be represented as 
\begin{equation}
S_{t'} = S_{t} + Z_{t'}, \quad 
\hbox{$Z_{t'}\sim\mathcal{N}(0,\mathbb{E}[\tilde{W}_{t'}^{2}]
- \mathbb{E}[\tilde{W}_{t}^{2}])$.}
\end{equation}
It is straightforward to confirm $\mathbb{E}[(S_{t'}-X)^{2}]=
\mathbb{E}[\tilde{W}_{t'}^{2}]$ and 
$\mathbb{E}[(S_{t'}-X)(S_{t}-X)]=\mathbb{E}[\tilde{W}_{t}^{2}]$. 

This representation implies that $S_{t}$ is a sufficient statistic for 
estimation of $X$ based on both $S_{t'}$ and $S_{t}$. Thus, we have 
$\mathbb{E}[X|S_{t'},S_{t}]=\mathbb{E}[X|S_{t}]$. Using this identity, we find 
that (\ref{posterior_covariance}) reduces to  
\begin{IEEEeqnarray}{rl}
&C(S_{t'},S_{t};\mathbb{E}[\tilde{W}_{t'}^{2}], \mathbb{E}[\tilde{W}_{t}^{2}])
- C(S_{t},S_{t};\mathbb{E}[\tilde{W}_{t}^{2}], \mathbb{E}[\tilde{W}_{t}^{2}]) 
\nonumber \\
=& \{\mathbb{E}[X|S_{t}] - \mathbb{E}[X|S_{t'}]\}
\{\mathbb{E}[X| S_{t'}, S_{t}] - \mathbb{E}[X|S_{t}]\}=0.
\end{IEEEeqnarray}
Thus, the second property holds. 

Before proving the last property, we prove the monotonicity 
$\Delta\Sigma_{\tau,t}=[\boldsymbol{\Sigma}_{t}]_{\tau,\tau} 
- [\boldsymbol{\Sigma}_{t}]_{\tau+1,\tau+1}>0$ for all $\tau$.  
For that purpose, we evaluate the determinant $\det\boldsymbol{\Sigma}_{t}$. 
Subtracting the $(\tau+1)$th column in 
$\boldsymbol{\Sigma}_{t}$ from the $\tau$th column for $\tau=0,\ldots,t-1$, 
we use the assumptions $[\boldsymbol{\Sigma}_{t}]_{\tau',\tau}
=[\boldsymbol{\Sigma}_{t}]_{\tau,\tau'}
=[\boldsymbol{\Sigma}_{t}]_{\tau,\tau}$ to have 
\begin{IEEEeqnarray}{rl}
\det\boldsymbol{\Sigma}_{t} 
=& \begin{vmatrix}
\Delta\Sigma_{0,t} & \cdots & \Delta\Sigma_{t-1,t} & 
[\boldsymbol{\Sigma}_{t}]_{t,t} \\
0 & \ddots & \vdots & \vdots \\
\vdots & \ddots & \Delta\Sigma_{t-1,t} & \vdots \\
0 & \cdots & 0 & [\boldsymbol{\Sigma}_{t}]_{t,t}
\end{vmatrix} 
\nonumber \\
=& [\boldsymbol{\Sigma}_{t}]_{t,t}\prod_{\tau=0}^{t-1}\Delta\Sigma_{\tau,t}. 
\label{determinant}
\end{IEEEeqnarray}
Since $\boldsymbol{\Sigma}_{t}$ has been assumed to be positive definite, 
the determinants of all square upper-left submatrices in 
$\boldsymbol{\Sigma}_{t}$ have to be positive. From (\ref{determinant}) we 
arrive at $\Delta\Sigma_{\tau,t}>0$ for all $\tau\in\{0,\ldots,t-1\}$. 

Finally, we prove the last property. Using $\Delta\Sigma_{\tau,t}>0$, 
the AWGN measurements $\{Y_{\tau}\}_{\tau=0}^{t}$ can be represented as 
\begin{equation}
Y_{t} = X + V_{t}, 
\quad 
Y_{\tau-1} = Y_{\tau} + V_{\tau-1}
\end{equation}
for $\tau\in\{1,\ldots,t\}$, where $\{V_{\tau}\}$ are independent zero-mean 
Gaussian random variables with variance 
$\mathbb{E}[V_{t}^{2}]=[\boldsymbol{\Sigma}_{t}]_{t,t}$ and 
$\mathbb{E}[V_{\tau-1}^{2}]=\Delta\Sigma_{\tau-1,t}>0$. 
This representation implies that $Y_{t}$ is a sufficient statistic for 
estimation of $X$ based on the AWGN measurements $\{Y_{\tau}\}_{\tau=0}^{t}$. 
Thus, we have the identity 
$\mathbb{E}[\{X-f_{\mathrm{opt}}(S_{t};\mathbb{E}[\tilde{W}_{t}^{2}])\}^{2}]
=\mathbb{E}[\{X-f_{\mathrm{opt}}(Y_{t};[\boldsymbol{\Sigma}_{t}]_{t,t})\}^{2}]$, 
which is equivalent to $\mathbb{E}[\tilde{W}_{t}^{2}]
=[\boldsymbol{\Sigma}_{t}]_{t,t}$. 
\end{IEEEproof}

The first property in Lemma~\ref{lemma1} is used to prove that 
the mean-square error (MSE) for Bayes-optimal LM-OAMP is monotonically 
non-increasing. This property was utilized in convergence analysis for 
memory AMP~\cite{Liu21}. The remaining two properties are used to prove the 
reduction of Bayes-optimal LM-OAMP to Bayes-optimal OAMP. 

\section{Long-Memory OAMP} \label{sec3}
\subsection{Long-Memory Processing}
LM-OAMP is composed of two modules---called modules~A and~B. Module~A 
uses a linear filter to mitigate multiuser interference while module~B 
utilizes an element-wise nonlinear denoiser for signal reconstruction. 
A estimator of the signal vector $\boldsymbol{x}$ is computed via MP 
between the two modules. 

Each module employs LM processing, in which messages in all preceding 
iterations are used to update the current message, while messages only in the 
latest iteration are utilized in conventional MP. Let 
$\boldsymbol{x}_{\mathrm{A}\to\mathrm{B},t}\in\mathbb{R}^{N}$ and 
$\{v_{\mathrm{A}\to\mathrm{B},t',t}\in\mathbb{R}\}_{t'=0}^{t}$ denote messages that 
are passed from module~A to module~B in iteration~$t$. The former 
$\boldsymbol{x}_{\mathrm{A}\to\mathrm{B},t}$ is an estimator 
of $\boldsymbol{x}$ while the latter 
$v_{\mathrm{A}\to\mathrm{B},t',t}$ corresponds to an estimator for the error 
covariance $N^{-1}\mathbb{E}[
(\boldsymbol{x}_{\mathrm{A}\to\mathrm{B},t'}-\boldsymbol{x})^{\mathrm{T}}
(\boldsymbol{x}_{\mathrm{A}\to\mathrm{B},t}-\boldsymbol{x})]$. The messages 
used in iteration~$t$ are written as 
$\boldsymbol{X}_{\mathrm{A}\to\mathrm{B},t}=(\boldsymbol{x}_{\mathrm{A}\to\mathrm{B},0},
\ldots,\boldsymbol{x}_{\mathrm{A}\to\mathrm{B},t})\in\mathbb{R}^{N\times(t+1)}$ and 
symmetric $\boldsymbol{V}_{\mathrm{A}\to\mathrm{B},t}\in\mathbb{R}^{(t+1)\times(t+1)}$ 
with $[\boldsymbol{V}_{\mathrm{A}\to\mathrm{B},t}]_{\tau',\tau}
=v_{\mathrm{A}\to\mathrm{B},\tau',\tau}$ for all $\tau'\leq\tau$. 

Similarly, we define the corresponding messages passed from module~B to 
module~A in iteration~$t$ as $\boldsymbol{x}_{\mathrm{B}\to\mathrm{A},t}
\in\mathbb{R}^{N}$ and $\{v_{\mathrm{B}\to\mathrm{A},t',t}\in\mathbb{R}\}_{t'=0}^{t}$.  
They are compactly written as $\boldsymbol{X}_{\mathrm{B}\to\mathrm{A},t}
=(\boldsymbol{x}_{\mathrm{B}\to\mathrm{A},0},
\ldots,\boldsymbol{x}_{\mathrm{B}\to\mathrm{A},t})\in\mathbb{R}^{N\times(t+1)}$ and 
symmetric $\boldsymbol{V}_{\mathrm{B}\to\mathrm{A},t}\in\mathbb{R}^{(t+1)\times(t+1)}$ 
with $[\boldsymbol{V}_{\mathrm{B}\to\mathrm{A},t}]_{\tau',\tau}
=v_{\mathrm{B}\to\mathrm{A},\tau',\tau}$ for all $\tau'\leq\tau$. 

Asymptotic Gaussianity for estimation errors is postulated in formulating 
LM-OAMP. While asymptotic Gaussianity is defined and proved shortly, a 
rough interpretation is that estimation errors are jointly 
Gaussian-distributed in the large system limit, i.e.\  
$(\boldsymbol{x}_{\mathrm{A}\to\mathrm{B},t}-\boldsymbol{x})
(\boldsymbol{x}_{\mathrm{A}\to\mathrm{B},t'}-\boldsymbol{x})^{\mathrm{T}}\sim
\mathcal{N}(\boldsymbol{0},v_{\mathrm{A}\to\mathrm{B},t',t}\boldsymbol{I}_{N})$  
and $(\boldsymbol{x}_{\mathrm{B}\to\mathrm{A},t}-\boldsymbol{x})
(\boldsymbol{x}_{\mathrm{B}\to\mathrm{A},t'}-\boldsymbol{x})^{\mathrm{T}}\sim
\mathcal{N}(\boldsymbol{0},v_{\mathrm{B}\to\mathrm{A},t',t}\boldsymbol{I}_{N})$. 
This rough interpretation is too strong to justify. Nonetheless, it helps us 
understand update rules in LM-OAMP.  

\subsection{Module A (Linear Estimation)}
Module~A utilizes $\boldsymbol{X}_{\mathrm{B}\to\mathrm{A},t}$ and 
$\boldsymbol{V}_{\mathrm{B}\to\mathrm{A},t}$ provided by module~B to compute the 
mean and covariance messages $\boldsymbol{x}_{\mathrm{A}\to\mathrm{B},t}$ and 
$\{v_{\mathrm{A}\to\mathrm{B},t',t}\}_{t'=0}^{t}$ in iteration~$t$. A first step is 
computation of a sufficient statistic for estimation of $\boldsymbol{x}$. 
According to (\ref{sufficient_statistic}) and (\ref{sufficient_covariance}), 
we define a sufficient statistic 
$\boldsymbol{x}_{\mathrm{B}\to \mathrm{A},t}^{\mathrm{suf}}$ and the corresponding 
covariance $v_{\mathrm{B}\to \mathrm{A},t',t}^{\mathrm{suf}}$ as   
\begin{equation} \label{statistic_mean_A}
\boldsymbol{x}_{\mathrm{B}\to \mathrm{A},t}^{\mathrm{suf}} 
= \frac{\boldsymbol{X}_{\mathrm{B}\to \mathrm{A},t}
\boldsymbol{V}_{\mathrm{B}\to \mathrm{A},t}^{-1}\boldsymbol{1}}
{\boldsymbol{1}^{\mathrm{T}}\boldsymbol{V}_{\mathrm{B}\to \mathrm{A},t}^{-1}
\boldsymbol{1}},  
\end{equation} 
\begin{equation} \label{statistic_covariance_A} 
v_{\mathrm{B}\to \mathrm{A},t',t}^{\mathrm{suf}}
= \frac{1}
{\boldsymbol{1}^{\mathrm{T}}\boldsymbol{V}_{\mathrm{B}\to \mathrm{A},t}^{-1}
\boldsymbol{1}}
\quad \hbox{for all $t'\leq t$.} 
\end{equation}
In the initial iteration $t=0$, the initial values 
$\boldsymbol{x}_{\mathrm{B}\to\mathrm{A},0}=\boldsymbol{0}$ and 
$v_{\mathrm{B}\to\mathrm{A},0,0}=\mathbb{E}[\|\boldsymbol{x}\|^{2}]/N$ are used 
to compute (\ref{statistic_mean_A}) and (\ref{statistic_covariance_A}). 

The sufficient statistic~(\ref{statistic_mean_A}) is equivalent to optimized 
LM damping of all preceding messages 
$\{\boldsymbol{x}_{\mathrm{B}\to\mathrm{A},t'}\}_{t'=0}^{t}$ in \cite{Liu21}, which 
was obtained as a solution to an optimization problem based on state 
evolution results. However, this statistical interpretation is a key 
technical tool in proving the main theorem.   

A second step is computation of posterior mean 
$\boldsymbol{x}_{\mathrm{A},t}^{\mathrm{post}}\in\mathbb{R}^{N}$ and 
covariance $\{v_{\mathrm{A},t',t}^{\mathrm{post}}\}_{t'=0}^{t}$. A linear filter 
$\boldsymbol{W}_{t}\in\mathbb{R}^{M\times N}$ is used to obtain    
\begin{equation} \label{post_mean_A}
\boldsymbol{x}_{\mathrm{A},t}^{\mathrm{post}} 
= \boldsymbol{x}_{\mathrm{B}\to \mathrm{A},t}^{\mathrm{suf}} 
+ \boldsymbol{W}_{t}^{\mathrm{T}}(\boldsymbol{y} 
- \boldsymbol{A}\boldsymbol{x}_{\mathrm{B}\to \mathrm{A},t}^{\mathrm{suf}}), 
\end{equation}
\begin{equation} \label{post_covariance_A}
v_{\mathrm{A},t',t}^{\mathrm{post}} 
= \gamma_{t',t}v_{\mathrm{B}\to \mathrm{A},t',t}^{\mathrm{suf}}
+ \frac{\sigma^{2}}{N}\mathrm{Tr}\left(
 \boldsymbol{W}_{t'}\boldsymbol{W}_{t}^{\mathrm{T}}
\right),
\end{equation}
with 
\begin{equation} \label{gamma_t't}
\gamma_{t',t} 
= \frac{1}{N}\mathrm{Tr}\left\{
 \left(
  \boldsymbol{I}_{N} - \boldsymbol{W}_{t'}^{\mathrm{T}}\boldsymbol{A}
 \right)^{\mathrm{T}}
 \left(
  \boldsymbol{I}_{N} - \boldsymbol{W}_{t}^{\mathrm{T}}\boldsymbol{A}
 \right)
\right\}. 
\end{equation}

In this paper, we focus on the linear minimum mean-square error (LMMSE) filter 
\begin{equation} \label{LMMSE}
\boldsymbol{W}_{t} 
= v_{\mathrm{B}\to \mathrm{A},t,t}^{\mathrm{suf}}\left(
 \sigma^{2}\boldsymbol{I}_{M} + v_{\mathrm{B}\to \mathrm{A},t,t}^{\mathrm{suf}}
 \boldsymbol{A}\boldsymbol{A}^{\mathrm{T}}
\right)^{-1}\boldsymbol{A}. 
\end{equation}
The LMMSE filter minimizes the posterior variance 
$v_{\mathrm{A},t,t}^{\mathrm{post}}$ among all possible linear filters.  

The last step is computation of extrinsic messages 
$\boldsymbol{x}_{\mathrm{A}\to\mathrm{B},t}\in\mathbb{R}^{N}$ and 
$\{v_{\mathrm{A}\to\mathrm{B},t',t}\}_{t'=0}^{t}$ to realize asymptotic 
Gaussianity in module~B. Let 
\begin{equation} \label{xi_A_t't} 
\xi_{\mathrm{A},t',t} 
= \xi_{\mathrm{A},t}\frac{\boldsymbol{e}_{t'}^{\mathrm{T}}
\boldsymbol{V}_{\mathrm{B}\to \mathrm{A},t}^{-1}\boldsymbol{1}}
{\boldsymbol{1}^{\mathrm{T}}\boldsymbol{V}_{\mathrm{B}\to \mathrm{A},t}^{-1}
\boldsymbol{1}}, 
\end{equation}
where $\boldsymbol{e}_{t}$ is the $t$th column of $\boldsymbol{I}$, 
with
\begin{equation} \label{xi_A_t}
\xi_{\mathrm{A},t} 
= \frac{1}{N}\mathrm{Tr}\left(
 \boldsymbol{I}_{N} - \boldsymbol{W}_{t}^{\mathrm{T}}\boldsymbol{A}
\right). 
\end{equation}
The extrinsic mean $\boldsymbol{x}_{\mathrm{A}\to\mathrm{B},t}$ and covariance 
$\{v_{\mathrm{A}\to\mathrm{B},t',t}\}_{t'=0}^{t}$ are computed as  
\begin{equation} \label{mean_AB} 
\boldsymbol{x}_{\mathrm{A}\to\mathrm{B},t} 
= \frac{\boldsymbol{x}_{\mathrm{A},t}^{\mathrm{post}} 
- \sum_{t'=0}^{t}\xi_{\mathrm{A},t',t}
\boldsymbol{x}_{\mathrm{B}\to \mathrm{A},t'}}
{1 - \xi_{\mathrm{A},t}},
\end{equation}
\begin{equation} \label{covariance_AB}
v_{\mathrm{A}\to\mathrm{B},t',t}
= \frac{v_{\mathrm{A},t',t}^{\mathrm{post}} 
- \xi_{\mathrm{A},t'}\xi_{\mathrm{A},t}v_{\mathrm{B}\to \mathrm{A},t',t}^{\mathrm{suf}}}
{(1-\xi_{\mathrm{A},t'})(1-\xi_{\mathrm{A},t})}.  
\end{equation}

The numerator in (\ref{mean_AB}) is the so-called Onsager correction of the 
posterior mean $\boldsymbol{x}_{\mathrm{A},t}^{\mathrm{post}}$ to realize 
asymptotic Gaussianity. The denominator can be set to an arbitrary constant. 
In this paper, we set the denominator so as to minimize the extrinsic variance 
$v_{\mathrm{A}\to\mathrm{B},t,t}$ for the LMMSE filter~(\ref{LMMSE}). 
See \cite[Section~II]{Takeuchi213} for the details. 

\subsection{Module B (Nonlinear Estimation)}
Module~B uses $\boldsymbol{X}_{\mathrm{A}\to\mathrm{B},t}$ and 
$\boldsymbol{V}_{\mathrm{A}\to\mathrm{B},t}$ to compute the messages 
$\boldsymbol{x}_{\mathrm{B}\to\mathrm{A},t+1}$ and 
$\{v_{\mathrm{B}\to\mathrm{A},t',t+1}\}_{t'=0}^{t+1}$ in the same manner as in module~A. 
A sufficient statistic $\boldsymbol{x}_{\mathrm{A}\to \mathrm{B},t}^{\mathrm{suf}}
\in\mathbb{R}^{N}$ and the corresponding covariance 
$\{v_{\mathrm{A}\to \mathrm{B},t',t}^{\mathrm{suf}}\in\mathbb{R}\}_{t'=0}^{t}$ are 
computed as   
\begin{equation} \label{statistic_mean_B} 
\boldsymbol{x}_{\mathrm{A}\to \mathrm{B},t}^{\mathrm{suf}} 
= \frac{\boldsymbol{X}_{\mathrm{A}\to \mathrm{B},t}
\boldsymbol{V}_{\mathrm{A}\to \mathrm{B},t}^{-1}\boldsymbol{1}}
{\boldsymbol{1}^{\mathrm{T}}
\boldsymbol{V}_{\mathrm{A}\to \mathrm{B},t}^{-1}\boldsymbol{1}},
\end{equation}
\begin{equation}
v_{\mathrm{A}\to \mathrm{B},t',t}^{\mathrm{suf}}
= \frac{1}{\boldsymbol{1}^{\mathrm{T}}\boldsymbol{V}_{\mathrm{A}\to \mathrm{B},t}^{-1}
\boldsymbol{1}}
\quad \hbox{for all $t'\leq t$.} 
\end{equation}

Module~B next computes the posterior messages 
$\boldsymbol{x}_{\mathrm{B},t+1}^{\mathrm{post}}$ and  
$\{v_{\mathrm{B},t'+1,t+1}^{\mathrm{post}}\}_{t'=0}^{t}$ with the posterior mean 
$f_{\mathrm{opt}}(\cdot;\cdot)$ and covariance~(\ref{posterior_covariance}) 
for the correlated AWGN measurements,   
\begin{equation} \label{post_mean_B}
\boldsymbol{x}_{\mathrm{B},t+1}^{\mathrm{post}}
= f_{\mathrm{opt}}(\boldsymbol{x}_{\mathrm{A}\to\mathrm{B},t}^{\mathrm{suf}};
v_{\mathrm{A}\to \mathrm{B},t,t}^{\mathrm{suf}}),
\end{equation}
\begin{IEEEeqnarray}{r}
v_{\mathrm{B},t'+1,t+1}^{\mathrm{post}}
= \frac{1}{N}\sum_{n=1}^{N}
C\left(
 [\boldsymbol{x}_{\mathrm{A}\to\mathrm{B},t'}^{\mathrm{suf}}]_{n},
 [\boldsymbol{x}_{\mathrm{A}\to\mathrm{B},t}^{\mathrm{suf}}]_{n};
\right.
\nonumber \\
\left. 
 v_{\mathrm{A}\to \mathrm{B},t',t'}^{\mathrm{suf}}, 
 v_{\mathrm{A}\to \mathrm{B},t,t}^{\mathrm{suf}} 
\right), \label{post_covariance_B}
\end{IEEEeqnarray}
where the right-hand side (RHS) of (\ref{post_mean_B}) means the 
element-wise application of $f_{\mathrm{opt}}$ to 
$\boldsymbol{x}_{\mathrm{A}\to\mathrm{B},t}^{\mathrm{suf}}$. 
The posterior mean~(\ref{post_mean_B}) is used as an estimator of 
$\boldsymbol{x}$. 

To realize asymptotic Gaussianity in module~A, the extrinsic mean 
$\boldsymbol{x}_{\mathrm{B}\to\mathrm{A},t+1}$ 
and covariance $\{v_{\mathrm{B}\to\mathrm{A},t',t+1}\}_{t'=0}^{t+1}$ are fed back to 
module~A. Let 
\begin{equation} \label{xi_B_t't} 
\xi_{\mathrm{B},t',t} 
= \xi_{\mathrm{B},t}
\frac{\boldsymbol{e}_{t'}^{\mathrm{T}}
\boldsymbol{V}_{\mathrm{A}\to \mathrm{B},t}^{-1}\boldsymbol{1}}
{\boldsymbol{1}^{\mathrm{T}}\boldsymbol{V}_{\mathrm{A}\to \mathrm{B},t}^{-1}
\boldsymbol{1}},  
\end{equation}
with
\begin{equation} \label{xi_B_t}
\xi_{\mathrm{B},t} 
= \frac{1}{N}\sum_{n=1}^{N}f_{\mathrm{opt}}'
([\boldsymbol{x}_{\mathrm{A}\to \mathrm{B},t}^{\mathrm{suf}}]_{n};
v_{\mathrm{A}\to \mathrm{B},t}^{\mathrm{suf}}),   
\end{equation}
where the derivative is taken with respect to the first variable. 
The extrinsic messages are computed as 
\begin{equation} \label{mean_BA} 
\boldsymbol{x}_{\mathrm{B}\to\mathrm{A},t+1}
= \frac{\boldsymbol{x}_{\mathrm{B},t+1}^{\mathrm{post}} 
- \sum_{t'=0}^{t}\xi_{\mathrm{B},t',t}
\boldsymbol{x}_{\mathrm{A}\to \mathrm{B},t'}}{1 - \xi_{\mathrm{B},t}}, 
\end{equation}
\begin{equation} \label{covariance_BA}
v_{\mathrm{B}\to\mathrm{A},t'+1,t+1} 
= \frac{ v_{\mathrm{B},t'+1,t+1}^{\mathrm{post}}
- \xi_{\mathrm{B},t'}\xi_{\mathrm{B},t}v_{\mathrm{A}\to \mathrm{B},t',t}^{\mathrm{suf}}}
{(1-\xi_{\mathrm{B},t'})(1-\xi_{\mathrm{B},t})}
\end{equation}
for $t'\in\{0,\ldots,t\}$, with $v_{\mathrm{B}\to \mathrm{A},0,t+1}
=v_{\mathrm{B},t+1,t+1}^{\mathrm{post}}/(1 - \xi_{\mathrm{B},t})$. 

\section{Main Results} \label{sec4}
\subsection{State Evolution} 
The dynamics of LM-OAMP is analyzed via state evolution~\cite{Takeuchi211} 
in the large system limit. Asymptotic Gaussianity has been proved for a 
general error model proposed in \cite{Takeuchi211}. Thus, the main part in 
state evolution analysis is to prove the inclusion of the error model for 
LM-OAMP into the general error model.  

Before presenting state evolution results, we first summarize technical 
assumptions. 
\begin{assumption} \label{assumption_x}
The signal vector $\boldsymbol{x}$ has i.i.d.\ elements with zero mean, 
unit variance, and bounded $(2+\epsilon)$th moment for some $\epsilon>0$. 
\end{assumption}

Assumption~\ref{assumption_x} is a simplifying assumption. To relax 
Assumption~\ref{assumption_x}, we need non-separable 
denoisers~\cite{Berthier19,Ma19,Fletcher19}. 
\begin{assumption} \label{assumption_A}
The sensing matrix $\boldsymbol{A}$ is right-orthogonally invariant: 
For any orthogonal matrix $\boldsymbol{\Phi}$ independent of $\boldsymbol{A}$, 
the equivalence in distribution 
$\boldsymbol{A}\boldsymbol{\Phi}\sim\boldsymbol{A}$ holds. 
More precisely, in the singular-value decomposition (SVD)  
$\boldsymbol{A}=\boldsymbol{U}\boldsymbol{\Sigma}
\boldsymbol{V}^{\mathrm{T}}$ the orthogonal matrix $\boldsymbol{V}$ is 
independent of $\boldsymbol{U}\boldsymbol{\Sigma}$ and 
Haar-distributed~\cite{Tulino04,Hiai00}. Furthermore, 
the empirical eigenvalue distribution of $\boldsymbol{A}^{\mathrm{T}}
\boldsymbol{A}$ converges almost surely to a compactly supported deterministic 
distribution with unit first moment in the large system limit. 
\end{assumption} 

The right-orthogonal invariance is a key assumption in state evolution 
analysis. The unit-first-moment assumption implies the almost sure convergence 
$N^{-1}\mathrm{Tr}(\boldsymbol{A}^{\mathrm{T}}\boldsymbol{A})\ato1$. 

\begin{assumption} \label{assumption_denoiser}
The Bayes-optimal denoiser $f_{\mathrm{opt}}$ in module~B is 
nonlinear and Lipschitz-continuous. 
\end{assumption}

The nonlinearity is required to guarantee the asymptotic positive definiteness 
of $\boldsymbol{V}_{\mathrm{A}\to\mathrm{B},t}$ and 
$\boldsymbol{V}_{\mathrm{B}\to\mathrm{A},t}$. 
It is an interesting open question whether any Bayes-optimal denoiser is 
Lipschitz-continuous under Assumption~\ref{assumption_x}. 

We next define state evolution recursions for LM-OAMP, which are 2D 
discrete systems with respect to two positive-definite symmetric matrices 
$\bar{\boldsymbol{V}}_{\mathrm{A}\to\mathrm{B},t}\in\mathbb{R}^{(t+1)\times(t+1)}$ and 
$\bar{\boldsymbol{V}}_{\mathrm{B}\to\mathrm{A},t}\in\mathbb{R}^{(t+1)\times(t+1)}$. 
We write the $(\tau',\tau)$ elements of 
$\bar{\boldsymbol{V}}_{\mathrm{A}\to\mathrm{B},t}\in\mathbb{R}^{(t+1)\times(t+1)}$ 
and $\bar{\boldsymbol{V}}_{\mathrm{B}\to\mathrm{A},t}\in\mathbb{R}^{(t+1)\times(t+1)}$ 
as $\bar{v}_{\mathrm{A}\to\mathrm{B},\tau',\tau}$ and 
$\bar{v}_{\mathrm{B}\to\mathrm{A},\tau',\tau}$ for $\tau',\tau\in\{0,\ldots,t\}$, 
respectively. 

Consider the initial condition $\bar{v}_{\mathrm{B}\to \mathrm{A},0,0}=1$. 
State evolution recursions for module~A are given by   
\begin{equation} \label{SE_statistic_A}
\bar{v}_{\mathrm{B}\to \mathrm{A},t',t}^{\mathrm{suf}} 
= \frac{1}{\boldsymbol{1}^{\mathrm{T}}
\bar{\boldsymbol{V}}_{\mathrm{B}\to \mathrm{A},t}^{-1}\boldsymbol{1}}
\quad \hbox{for all $t'\leq t$},  
\end{equation}
\begin{equation} \label{SE_post_A} 
\bar{v}_{\mathrm{A},t',t}^{\mathrm{post}} 
= \lim_{M=\delta N\to\infty}\left\{
 \gamma_{t',t}\bar{v}_{\mathrm{B}\to \mathrm{A},t',t}^{\mathrm{suf}}
 + \frac{\sigma^{2}}{N}\mathrm{Tr}\left(
  \boldsymbol{W}_{t'}\boldsymbol{W}_{t}^{\mathrm{T}}
 \right)
\right\},
\end{equation}
\begin{equation} \label{SE_AB}
\bar{v}_{\mathrm{A}\to \mathrm{B},t',t} 
= \frac{\bar{v}_{\mathrm{A},t',t}^{\mathrm{post}} 
- \bar{\xi}_{\mathrm{A},t'}\bar{\xi}_{\mathrm{A},t}
\bar{v}_{\mathrm{B}\to \mathrm{A},t',t}^{\mathrm{suf}}}
{(1-\bar{\xi}_{\mathrm{A},t'})(1-\bar{\xi}_{\mathrm{A},t})},
\end{equation}
with $\bar{\xi}_{\mathrm{A},t} = \bar{v}_{\mathrm{A},t,t}^{\mathrm{post}}
/\bar{v}_{\mathrm{B}\to\mathrm{A},t,t}^{\mathrm{suf}}$. In (\ref{SE_post_A}), 
$\gamma_{t',t}$ is given by (\ref{gamma_t't}). The LMMSE filter 
$\boldsymbol{W}_{t}$ is defined as (\ref{LMMSE}) with 
$v_{\mathrm{B}\to\mathrm{A},t,t}^{\mathrm{suf}}$ replaced by 
$\bar{v}_{\mathrm{B}\to\mathrm{A},t,t}^{\mathrm{suf}}$. 

State evolution recursions for module~B are given by 
\begin{equation} \label{SE_statistic_B}
\bar{v}_{\mathrm{A}\to \mathrm{B},t',t}^{\mathrm{suf}}
= \frac{1}
{\boldsymbol{1}^{\mathrm{T}}
\bar{\boldsymbol{V}}_{\mathrm{A}\to \mathrm{B},t,t}^{-1}\boldsymbol{1}} 
\quad \hbox{for $t'\leq t$,}  
\end{equation}
\begin{IEEEeqnarray}{rl}
\bar{v}_{\mathrm{B},t'+1,t+1}^{\mathrm{post}} 
= \mathbb{E}[&\{f_{\mathrm{opt}}(x_{1} + z_{t'};
\bar{v}_{\mathrm{A}\to\mathrm{B},t',t'}^{\mathrm{suf}}) - x_{1}\}
\nonumber \\
&\cdot\{f_{\mathrm{opt}}(x_{1} + z_{t};
\bar{v}_{\mathrm{A}\to\mathrm{B},t,t}^{\mathrm{suf}}) - x_{1}\}], 
\label{SE_post_B}
\end{IEEEeqnarray}
\begin{equation} \label{SE_BA}
\bar{v}_{\mathrm{B}\to \mathrm{A},t'+1,t+1} 
= \frac{\bar{v}_{\mathrm{B},t'+1,t+1}^{\mathrm{post}}
 - \bar{\xi}_{\mathrm{B},t'}\bar{\xi}_{\mathrm{B},t}
\bar{v}_{\mathrm{A}\to \mathrm{B},t',t}^{\mathrm{suf}}}
{(1-\bar{\xi}_{\mathrm{B},t'})(1-\bar{\xi}_{\mathrm{B},t})}   
\end{equation}
for $t', t\geq0$, with $\bar{v}_{\mathrm{B}\to \mathrm{A},0,t+1}
=\bar{v}_{\mathrm{B},t+1,t+1}^{\mathrm{post}}/(1 - \bar{\xi}_{\mathrm{B},t})$ and 
$\bar{\xi}_{\mathrm{B},t} = \bar{v}_{\mathrm{B},t+1,t+1}^{\mathrm{post}}
/\bar{v}_{\mathrm{A}\to\mathrm{B},t,t}^{\mathrm{suf}}$. In (\ref{SE_post_B}), 
$\{z_{\tau}\}$ are independent of the signal element $x_{1}$ 
and zero-mean Gaussian random variables with covariance 
$\mathbb{E}[z_{\tau'}z_{\tau}]=\bar{v}_{\mathrm{A}\to \mathrm{B},\tau',\tau}^{\mathrm{suf}}$. 

\begin{theorem} \label{theorem_SE}
Suppose that Assumptions~\ref{assumption_x}--\ref{assumption_denoiser} hold. 
Then, the covariance $N^{-1}(\boldsymbol{x}_{\mathrm{A},t'}^{\mathrm{post}}
-\boldsymbol{x})^{\mathrm{T}}
(\boldsymbol{x}_{\mathrm{A},t}^{\mathrm{post}}-\boldsymbol{x})$ and 
$N^{-1}(\boldsymbol{x}_{\mathrm{B},t'+1}^{\mathrm{post}}
-\boldsymbol{x})^{\mathrm{T}}
(\boldsymbol{x}_{\mathrm{B},t+1}^{\mathrm{post}}-\boldsymbol{x})$ 
converges almost surely to $\bar{v}_{\mathrm{A},t',t}^{\mathrm{post}}$
and $\bar{v}_{\mathrm{B},t'+1,t+1}^{\mathrm{post}}$ in the large system limit, 
where $\bar{v}_{\mathrm{A},t',t}^{\mathrm{post}}$
and $\bar{v}_{\mathrm{B},t'+1,t+1}^{\mathrm{post}}$ satisfy the state evolution 
recursions~(\ref{SE_statistic_A})--(\ref{SE_BA}). 
\end{theorem}
\begin{IEEEproof}
See \cite[Theorems~1 and~2]{Takeuchi213}. 
\end{IEEEproof}

The almost sure convergence $N^{-1}(\boldsymbol{x}_{\mathrm{B},t'+1}^{\mathrm{post}}
-\boldsymbol{x})^{\mathrm{T}}
(\boldsymbol{x}_{\mathrm{B},t+1}^{\mathrm{post}}-\boldsymbol{x})
\ato\bar{v}_{\mathrm{B},t'+1,t+1}^{\mathrm{post}}$ is the precise meaning of 
asymptotic Gaussianity. As shown in (\ref{SE_post_B}), 
$\bar{v}_{\mathrm{B},t'+1,t+1}^{\mathrm{post}}$ is given via the error covariance 
for the Bayes-optimal estimation of $x_{1}$ based on the correlated AWGN 
measurements $x_{1}+z_{t'}$ and $x_{1}+z_{t}$.  

Theorem~\ref{theorem_SE} implies that the covariance messages 
$v_{\mathrm{A},t',t}^{\mathrm{post}}$ and $v_{\mathrm{B},t'+1,t+1}^{\mathrm{post}}$
in LM-OAMP are consistent estimators for the covariance 
$N^{-1}(\boldsymbol{x}_{\mathrm{A},t'}^{\mathrm{post}}
-\boldsymbol{x})^{\mathrm{T}}
(\boldsymbol{x}_{\mathrm{A},t}^{\mathrm{post}}-\boldsymbol{x})$ and 
$N^{-1}(\boldsymbol{x}_{\mathrm{B},t'+1}^{\mathrm{post}}
-\boldsymbol{x})^{\mathrm{T}}
(\boldsymbol{x}_{\mathrm{B},t+1}^{\mathrm{post}}-\boldsymbol{x})$ 
in the large system limit, respectively. 

\subsection{Convergence} 
We next prove that the state evolution 
recursions~(\ref{SE_statistic_A})--(\ref{SE_BA}) for Bayes-optimal LM-OAMP 
is equivalent to those for conventional Bayes-optimal OAMP and that they 
converge to a fixed point, i.e.\ $\bar{v}_{\mathrm{B},t',t}^{\mathrm{post}}$ 
converges to a constant as $t$ tends to infinity for all $t'\leq t$. 
Note that, in general, the convergence of the diagonal element 
$\bar{v}_{\mathrm{B},t,t}^{\mathrm{post}}$ 
does not necessarily imply the convergence of the non-diagonal 
elements~\cite{Takeuchi22}. 

Before analyzing the convergence of the state evolution recursions, we review 
state evolution recursions for conventional Bayes-optimal 
OAMP~\cite{Ma17,Rangan192,Takeuchi201}.
\begin{IEEEeqnarray}{rl}
\bar{v}_{\mathrm{A},t}^{\mathrm{post}}
= \bar{v}_{\mathrm{B}\to \mathrm{A},t}  
- &\lim_{M=\delta N\to\infty}\frac{\bar{v}_{\mathrm{B}\to \mathrm{A},t}^{2}}{N}
\mathrm{Tr}\left\{
 \boldsymbol{A}\boldsymbol{A}^{\mathrm{T}}
\right.
\nonumber \\
&\left.
 \cdot\left(
  \sigma^{2}\boldsymbol{I}_{M} + \bar{v}_{\mathrm{B}\to \mathrm{A},t}\boldsymbol{A}
  \boldsymbol{A}^{\mathrm{T}}
 \right)^{-1}
\right\}, \label{OAMP_post_A}
\end{IEEEeqnarray} 
\begin{equation} \label{OAMP_AB}
\bar{v}_{\mathrm{A}\to\mathrm{B},t} 
= \left(
 \frac{1}{\bar{v}_{\mathrm{A},t}^{\mathrm{post}}}
 - \frac{1}{\bar{v}_{\mathrm{B}\to\mathrm{A},t}} 
\right)^{-1}, 
\end{equation}
\begin{equation} \label{OAMP_post_B}
\bar{v}_{\mathrm{B},t+1}^{\mathrm{post}}
= \mathbb{E}\left[
 \{f_{\mathrm{opt}}(x_{1}+z_{t};\bar{v}_{\mathrm{A}\to\mathrm{B},t}) - x_{1}\}^{2}
\right], 
\end{equation}
\begin{equation} \label{OAMP_BA}
\bar{v}_{\mathrm{B}\to\mathrm{A},t+1} 
= \left(
 \frac{1}{\bar{v}_{\mathrm{B},t+1}^{\mathrm{post}}}
 - \frac{1}{\bar{v}_{\mathrm{A}\to\mathrm{B},t}} 
\right)^{-1}, 
\end{equation}
with the initial condition $\bar{v}_{\mathrm{B}\to\mathrm{A},0}=1$, where 
$z_{t}$ in (\ref{OAMP_post_B}) is independent of $x_{1}$ and follows 
the zero-mean Gaussian distribution with variance 
$\bar{v}_{\mathrm{A}\to\mathrm{B},t}$. The MSE for the 
OAMP estimator in iteration~$t$ was proved to converge almost surely to 
$\bar{v}_{\mathrm{B},t}$ in the large system limit~\cite{Rangan192,Takeuchi201}. 
 
\begin{theorem} \label{theorem_convergence}
The state evolution recursions~(\ref{SE_statistic_A})--(\ref{SE_BA}) for 
Bayes-optimal LM-OAMP are equivalent to 
those~(\ref{OAMP_post_A})--(\ref{OAMP_BA}) for Bayes-optimal OAMP, i.e.\ 
$v_{\mathrm{B},t,t}^{\mathrm{post}}=v_{\mathrm{B},t}^{\mathrm{post}}$ holds for any $t$. 
Furthermore, $v_{\mathrm{B},t',t}^{\mathrm{post}}$ converges to a constant as 
$t$ tends to infinity for all $t'\leq t$. 
\end{theorem}
\begin{IEEEproof}
We first evaluate the RHS of (\ref{SE_post_A}). Let 
\begin{equation} \label{Xi}
\boldsymbol{\Xi}_{t} 
= \bar{v}_{\mathrm{B}\to \mathrm{A},t,t}^{\mathrm{suf}}\left(
 \sigma^{2}\boldsymbol{I}_{M} + \bar{v}_{\mathrm{B}\to \mathrm{A},t,t}^{\mathrm{suf}}
 \boldsymbol{A}\boldsymbol{A}^{\mathrm{T}}
\right)^{-1}. 
\end{equation}
Using $v_{\mathrm{B}\to \mathrm{A},t',t}^{\mathrm{suf}}\ato
\bar{v}_{\mathrm{B}\to \mathrm{A},t',t}^{\mathrm{suf}}
=\bar{v}_{\mathrm{B}\to \mathrm{A},t,t}^{\mathrm{suf}}$ for all $t'\leq t$ obtained 
from (\ref{SE_statistic_A}), the definition of $\gamma_{t',t}$ in 
(\ref{gamma_t't}), and the LMMSE filter~(\ref{LMMSE}), we have  
\begin{IEEEeqnarray}{l}
\bar{v}_{\mathrm{A},t',t}^{\mathrm{post}} 
\aeq \frac{\bar{v}_{\mathrm{B}\to \mathrm{A},t,t}^{\mathrm{suf}}}{N}\left\{
 N - \mathrm{Tr}\left(
  \boldsymbol{\Xi}_{t'}\boldsymbol{A}\boldsymbol{A}^{\mathrm{T}} 
  + \boldsymbol{\Xi}_{t}\boldsymbol{A}\boldsymbol{A}^{\mathrm{T}}
 \right)
\right\}
\nonumber \\
+ \frac{1}{N}\mathrm{Tr}\left\{
 \boldsymbol{\Xi}_{t'}\boldsymbol{A}\boldsymbol{A}^{\mathrm{T}}
 \boldsymbol{\Xi}_{t}
 \left(
  \sigma^{2}\boldsymbol{I}
  + \bar{v}_{\mathrm{B}\to \mathrm{A},t,t}^{\mathrm{suf}}
 \boldsymbol{A}\boldsymbol{A}^{\mathrm{T}}
 \right)
\right\} + o(1) \nonumber \\
= \bar{v}_{\mathrm{B}\to \mathrm{A},t,t}^{\mathrm{suf}}
- \frac{\bar{v}_{\mathrm{B}\to \mathrm{A},t,t}^{\mathrm{suf}}}{N}\mathrm{Tr}\left(
 \boldsymbol{\Xi}_{t}\boldsymbol{A}\boldsymbol{A}^{\mathrm{T}}
\right) + o(1), \label{SE_post_A_tmp}
\end{IEEEeqnarray}
which is equal to the RHS of (\ref{OAMP_post_A}) with 
$\bar{v}_{\mathrm{B}\to\mathrm{A},t}$ replaced by 
$\bar{v}_{\mathrm{B}\to\mathrm{A},t,t}^{\mathrm{suf}}$.  
In the derivation of the second equality we have used (\ref{Xi}). 
Note that (\ref{SE_post_A_tmp}) implies 
$\bar{v}_{\mathrm{A},t',t}^{\mathrm{post}}=\bar{v}_{\mathrm{A},t,t}^{\mathrm{post}}$  
for all $t'\leq t$. 

We next evaluate the RHSs of (\ref{SE_AB}) and (\ref{SE_BA}). Applying 
$\bar{v}_{\mathrm{B}\to\mathrm{A},t',t}^{\mathrm{suf}}
=\bar{v}_{\mathrm{B}\to\mathrm{A},t,t}^{\mathrm{suf}}$ obtained from 
(\ref{SE_statistic_A}), $\bar{v}_{\mathrm{A},t',t}^{\mathrm{post}}
=\bar{v}_{\mathrm{A},t,t}^{\mathrm{post}}$, and 
$\bar{\xi}_{\mathrm{A},t}=\bar{v}_{\mathrm{A},t,t}^{\mathrm{post}}/
\bar{v}_{\mathrm{B}\to\mathrm{A},t,t}^{\mathrm{suf}}$ to (\ref{SE_AB}), we have 
\begin{equation}
\bar{v}_{\mathrm{A}\to \mathrm{B},t',t} 
= \left(
 \frac{1}{\bar{v}_{\mathrm{A},t,t}^{\mathrm{post}}}
 - \frac{1}{\bar{v}_{\mathrm{B}\to\mathrm{A},t,t}^{\mathrm{suf}}}
\right)^{-1}.
\end{equation}
Similarly, we use $\bar{v}_{\mathrm{B},t',t}^{\mathrm{post}}
=\bar{v}_{\mathrm{B},t,t}^{\mathrm{post}}$  for (\ref{SE_post_B}), 
obtained from the second property in Lemma~\ref{lemma1}, 
to find that the RHS of (\ref{SE_BA}) reduces to 
\begin{equation}
\bar{v}_{\mathrm{B}\to \mathrm{A},t'+1,t+1} 
= \left(
 \frac{1}{\bar{v}_{\mathrm{B},t+1,t+1}^{\mathrm{post}}}
 - \frac{1}{\bar{v}_{\mathrm{A}\to\mathrm{B},t,t}^{\mathrm{suf}}}
\right)^{-1}.
\end{equation}

To prove $\bar{v}_{\mathrm{B},t,t}^{\mathrm{post}}=\bar{v}_{\mathrm{B},t}^{\mathrm{post}}$ 
for any $t$, it is sufficient to show $\bar{v}_{\mathrm{A}\to\mathrm{B},t,t}
=\bar{v}_{\mathrm{A}\to\mathrm{B},t,t}^{\mathrm{suf}}$ and 
$\bar{v}_{\mathrm{B}\to\mathrm{A},t,t}=\bar{v}_{\mathrm{B}\to\mathrm{A},t,t}^{\mathrm{suf}}$. 
These identities follow immediately from the last property in 
Lemma~\ref{lemma1}. Thus, the state evolution recursions for Bayes-optimal 
LM-OAMP are equivalent to those for Bayes-optimal OAMP. 

Finally, we prove the convergence of the state evolution recursions for 
Bayes-optimal LM-OAMP. The first property in Lemma~\ref{lemma1} implies that 
$\{\bar{v}_{\mathrm{B},t,t}^{\mathrm{post}}\geq0\}$ is a monotonically non-increasing 
sequence as $t$ grows. Thus, $\bar{v}_{\mathrm{B},t,t}^{\mathrm{post}}$ converges 
to a constant $\bar{v}_{\mathrm{B}}^{\mathrm{post}}$ as $t$ tends to infinity. 
Since $\bar{v}_{\mathrm{B},t',t}^{\mathrm{post}}=\bar{v}_{\mathrm{B},t,t}^{\mathrm{post}}$ 
holds for all $t'\leq t$, the convergence of the diagonal element 
$\bar{v}_{\mathrm{B},t,t}^{\mathrm{post}}$ implies that of the non-diagonal 
elements $\{\bar{v}_{\mathrm{B},t',t}^{\mathrm{post}}\}$. 
\end{IEEEproof}

Theorem~\ref{theorem_convergence} implies that the state evolution 
recursions~(\ref{OAMP_post_A})--(\ref{OAMP_BA}) for Bayes-optimal OAMP 
converge to a fixed point as $t$ tends to infinity. Furthermore, the LM-MP 
proof strategy developed in this paper claims the optimality of Bayes-optimal 
OAMP in terms of the convergence speed among all possible LM-MP algorithms 
included into a unified framework in \cite{Takeuchi211}.  

\balance

\bibliographystyle{IEEEtran}
\bibliography{IEEEabrv,kt-isit2022}

\end{document}